\documentclass[
 prb,
 amsmath,amssymb,
preprint,
 showpacs,
 floatfix
]{revtex4-1}

\usepackage{graphicx}
\usepackage{dcolumn}  
\usepackage{bm}       

\newcommand{\C}{$\,^\circ$C}
\newcommand{\etal}[1]{{#1}~\emph{et al.}}

\newcommand{\fig}[1]{Fig.~\ref{#1}}

\begin{document}

\title{Determination of the graphene growth mode on SiC(0001) and
SiC(000\=1)}
\author{J.B. Hannon, M. Copel, and R.M. Tromp}

\affiliation{IBM Research Division, T.J. Watson Research Center,
               Yorktown Heights, NY 10598}
\date{\today}


\begin{abstract}
We have determined the growth mode of graphene on
SiC(0001) and SiC(000\=1)
using ultra-thin, isotopically-labeled Si$^{13}$C
`marker layers' grown epitaxially on the Si$^{12}$C surfaces.
Few-layer graphene overlayers
were formed via thermal decomposition at elevated temperature.
For both surface terminations (Si-face and C-face),
we find that the $^{13}$C  is
located mainly in the outermost graphene layers, indicating that,
during decomposition, new graphene layers form underneath existing ones.
\end{abstract}




\maketitle

Graphene has
attracted considerable attention, in part,  due to
potential applications
in electronics~\cite{Nov05,Zha05,Ber06}.
Several techniques have
been employed to synthesize graphene: mechanical 
exfoliation, chemical vapor deposition onto metal surfaces,
and sublimation of Si from SiC.  This latter 
technique is attractive for electronics application 
because the graphene is formed directly on an insulating
substrate, although many aspects of the formation
process
are poorly understood.
Graphene has been grown via decomposition
on several polytypes of SiC.  Attention has
mainly focussed on the (0001)
and (000\=1) surfaces of the 4H and 6H polytypes.  These
polytypes correspond to different stacking sequences of
hexagonal SiC bilayers. Within the bilayers,
Si and C atoms are not co-planar.
At the (0001) surface, the Si atoms are 
outermost, while at the (000\=1) surface, the C atoms are 
outermost.

Perhaps surprisingly, the growth of graphene on these two
surfaces is significantly different.  On SiC(0001), graphene
layers are epitaxial, forming a
well-ordered $6\sqrt{3}\times6\sqrt{3}$
superstructure~\cite{For98}.  Even thick graphene films exhibit an
epitaxial relationship to the underlying substrate.  In
contrast, graphene grown on
SiC(000\=1) is more randomly oriented, indicating
a much weaker substrate influence~\cite{Has06,Has08}.

These differences in graphene crystallography suggest that the
growth mode of graphene might be very different on these
surfaces.  On SiC(0001), the observed epitaxy has led to
speculation that graphene grows
``from the inside out''~\cite{Emt08}.
That is, strong coupling to the 
substrate induces epitaxy in the first graphene layer.  
The second graphene layer forms \emph{under} the first,
and is oriented due to coupling to the SiC. The second
layer displaces the first layer outward.  This process
continues as the film grows thicker, resulting in a crystalline
film in which the outmost layer is the first layer to form.
The more-random stacking of graphene on SiC(000\=1) 
makes it difficult to infer the growth mode, but might indicate
that it is substantially different from SiC(0001).

Here, we use isotopic labeling to directly measure the graphene
growth mode on both SiC(0001) and SiC(000\=1). 
We 
grew ultra-thin epitaxial SiC layers via
chemical vapor deposition using a mixture of
disilane and isotopically pure
$^{13}$C ethylene.  The thickness of the epitaxial layers was
4-5 bilayers, so that the carbon content was slightly more
that that of a single graphene layer.
We then formed graphene via
SiC decomposition at elevated
temperature~\cite{bom75,For98}.  After graphene
formation we used
medium-energy ion scattering
(MEIS)~\cite{van85} to measure the depth distributions of both
$^{12}$C and $^{13}$C.
If graphene grows from the inside out $^{13}$C will be
located predominantly at the surface.  Conversely, 
if new graphene layers form
on top of existing ones, $^{13}$C will be situated
underneath a $^{12}$C overlayer.
We found that for graphene grown
on both
SiC(0001) and SiC(000\=1), $^{13}$C remains largely at
the surface, showing directly that new graphene layers form under
existing layers.

Our approach is similar to
that used by \etal{Gusev} to study the oxidation
of Si(001) using $^{16}$O$_2$ and $^{18}$O$_2$ ~\cite{Gus95}.
In that work the authors carried out a two-step oxidation of
Si(100) at
elevated temperature, e.g.\ first using $^{18}$O$_2$, and
then using $^{16}$O$_2$.
They then used MEIS to measure the depth profile of  $^{18}$O
and $^{16}$O.  From this data they determined
the growth mode of SiO$_2$ on Si(100) for
different processing conditions.  For example, they 
found that at 900\C, SiO$_2$ growth beyond 4-5~nm occurs
mainly by diffusion of
oxygen from the 
gas phase through the oxide to the Si interface~\cite{Gus95}.

\begin{figure}[h]
\includegraphics[width=5in]{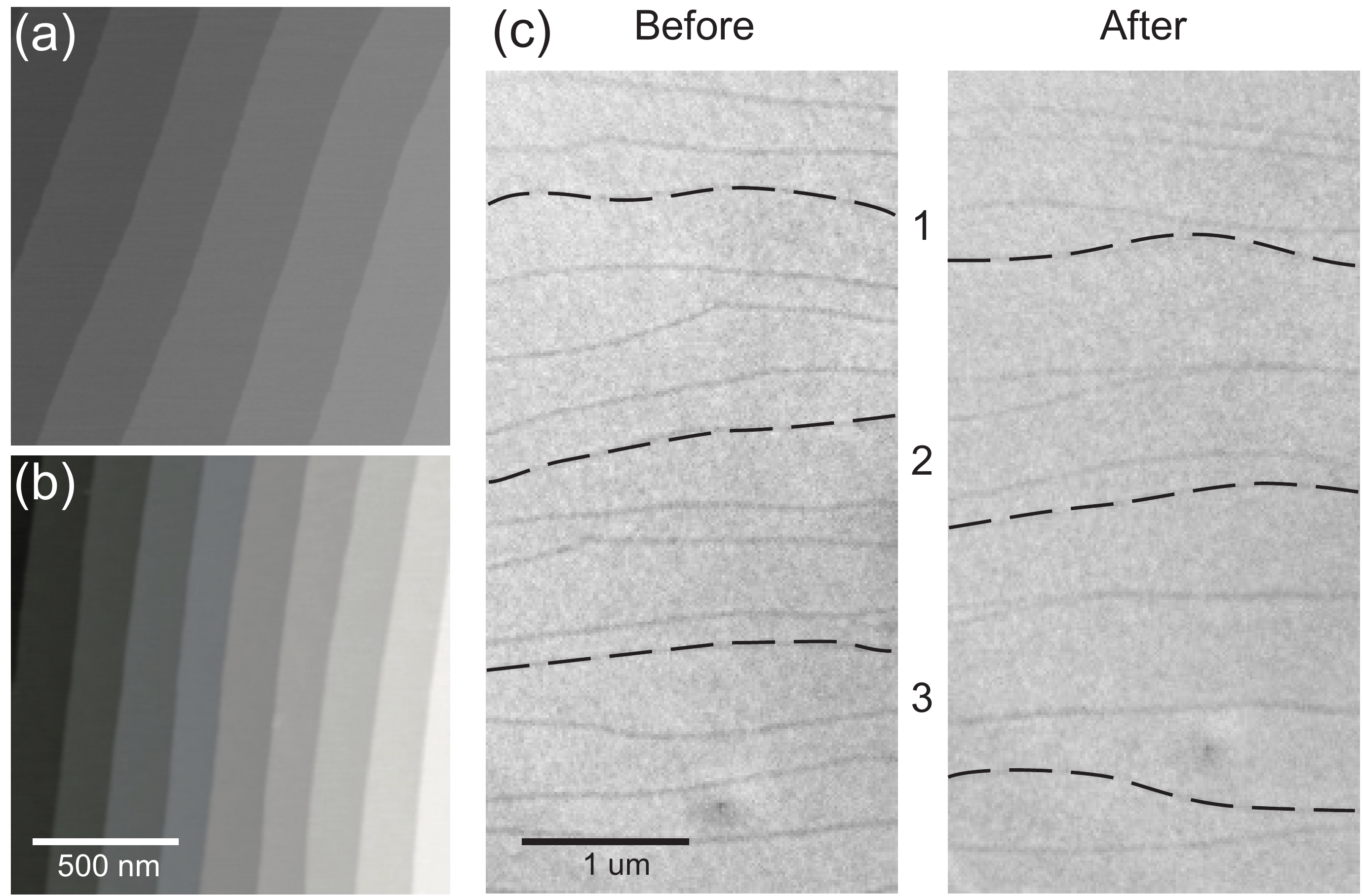}
\caption{AFM images recorded after annealing
 (a) SiC(0001)-6H, and (b) SiC(000\=1)-6H,
at 1200\C\ in 10$^{-6}$ Torr disilane.
A uniform step height of
0.8~nm is observed at both surfaces,
corresponding to three SiC bilayers. (c) Drift corrected 47~eV 
bright-field LEEM images recorded before and after
CVD growth of Si$^{13}$C.  The initial and final positions
of selected steps are shown, indicating downward step
flow during growth.}
\label{afm}
\end{figure}

SiC(0001)-6H and (000\=1)-6H surfaces were prepared by
degassing for several
hours in vacuum at
700\C.  Oxygen contamination was removed by annealing at 900\C\ in
a background pressure of $10^{-6}$~Torr dislane for 5 minutes.
After cleaning, the sample temperature was raised to 1200\C\ for 
10 minutes in the disilane.  As described
elsewhere~\cite{Tro09}, the
disilane prevents the formation of graphene.  Prolonged 
annealing at 1200\C\
leads to the formation of a uniform, reproducible surface
that consists of terraces 
bounded by straight steps with a uniform step
height of $\sim$0.8~nm (\fig{afm}).
This step height corresponds to three SiC bilayers.
%
%
%
%
%
For the Si-face, electron diffraction
analysis~\cite{Sun11} and first-principles
modeling~\cite{Han09} shows that triple-height steps are characteristic of the 
lowest-energy surface configuration for the 6H polytype.

Isotopically-labeled, epitaxial SiC
layers were then grown by exposing the clean surface
to a mixture of disilane (Si$_2$H$_6$) and isotopically-pure ethylene
($^{13}$C$_2$H$_4$).  In this way, SiC bilayers containing $^{13}$C,
i.e.\
Si$^{13}$C, were grown on top of the Si$^{12}$C substrate.  
Prior to ethylene exposure, the substrate temperature was
raised to 1200\C\ in 
$5\times10^{-6}$~Torr disilane.
The growth of Si$^{13}$C was initiated by adding $^{13}$C$_2$H$_4$
to the disilane until a total pressure of $7\times10^{-6}$~Torr was
achieved. The growth of epitaxial layers was monitored
\emph{in situ}
using LEEM.  Under these growth conditions, and 
for terraces widths of a few 100~nm, Si$^{13}$C grows via
step-flow, with three bilayers advancing simultaneously.  That is,
the step height of 0.76~nm is maintained.  During
 growth, steps advanced at a constant rate,
and the nucleation
of new SiC layers (e.g.\ islands) was not observed.
LEEM images recorded before and after Si$^{13}$C growth
are shown in \fig{afm}c.  The positions of selected steps are
marked
before and after growth, indicating that slightly
more than three bilayers of SiC were grown.

\begin{figure}[h]
\includegraphics[width=4in]{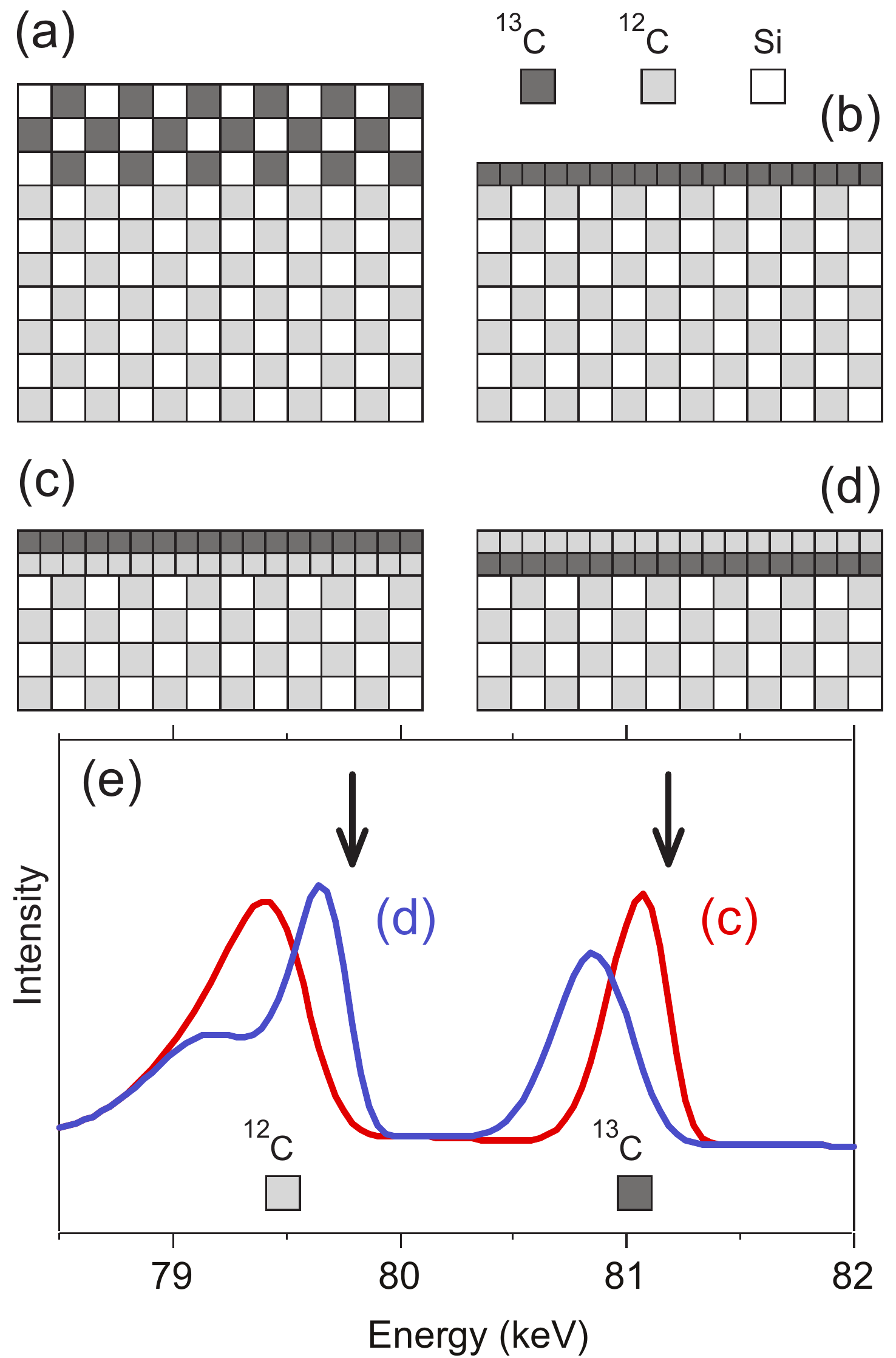}
\caption{(a) Three epitaxial Si$^{13}$C bilayers grown on
bulk Si$^{12}$C. (b) Resulting $^{13}$C graphene layer
that forms upon decomposition of three bilayers. (c,d)
two possible structures for bilayer graphene that forms
when further decomposition occurs.  (e) Simulated 100~keV
MEIS energy distributions for the structres shown in (c,d).
Arrows indicate the final energy of protons scattered from
$^{12}$C and $^{13}$C at the surface.
}
\label{model}
\end{figure}

The measured step velocity corresponded to a 
growth rate of approximately
one SiC bilayer per minute.
After the growth of about three SiC bilayers, the ethylene flow
was stopped,
but the disilane background pressure was maintained in order to
prevent graphene formation.  When the ethylene was removed, the
step motion ceased.  The structure of the resulting surface
is shown schematically in \fig{model}a.  Each bilayer consists of
a 50/50 mixture of Si and C.  At the (0001) surface, the Si atoms
are displaced outward relative to the C atoms.  The reverse is true
at the (000\=1) surface.
%
%
The structure shown in \fig{model}a corresponds to three
Si$^{13}$C bilayers grown on bulk Si$^{12}$C.  When
annealed above the Si sublimation temperature, graphene
will form.
The 6H polytype decomposes in units of three
bilayers~\cite{Han08}, and
the 
carbon density in graphene is equal to that of three
bilayers.  Therefore, the structure shown in \fig{model}b
-- a pure $^{13}$C graphene layer on bulk Si$^{12}$C --
should result if the decomposition is halted after 
the formation of a single graphene layer.  This `marker' layer
can be used to determine where additional graphene layers
form, giving insight into the kinetics of
graphene formation.  For example, if additional graphene layers
form under pre-existing layers (\fig{model}c), the  
$^{13}$C layer will always be outermost.  Conversely, if
additional layers form on top of pre-existing graphene, the 
$^{13}$C layer will be 
located at the graphene/SiC interface,
underneath a $^{12}$C graphene overlayer (\fig{model}d).

MEIS can be used to  distinguish between these two 
possible growth modes.  In our MEIS experiments, a
100~keV proton beam was incident normal to the surface and
the kinetic energy of the backscattered protons was
measured over a range of scattering angles near 110$^\circ$.
The energy of the backscattered protons can be used to 
determine the depth distribution and mass of the near-surface
atoms.  Two basic processes determine the final proton
energy.  First, when a proton scatters elastically from
a nucleus, conservation of energy and momentum leads to 
a characteristic backscatter energy given by the mass of the 
target nucleus and the scattering angle.  This dependence
can be used to measure the absolute concentrations of 
$^{12}$C and $^{13}$C. Each isotope gives rise to a
characteristic peak in the proton energy spectrum.
Second, as the proton travels through
the sample, inelastic electronic interactions give rise
to a characteristic energy loss per unit length travelled.
For most materials, the maximum
energy loss per unit length (for protons) occurs near 100~keV, which 
makes MEIS particularly surface sensitive.
Protons that scatter from nuclei located below the surface
will  have a lower kinetic energy than those that scatter
from nuclei at the surface.
The depth distribution will give
a characteristic shape and width to the peak in the 
proton energy spectrum.
These features of MEIS make it possible to
measure
accurate depth profiles for both $^{12}$C and $^{13}$C~\cite{Cop11}.

In \fig{model}e the calculated energy distribution of scattered
protons is shown
for the structures indicated in \fig{model}c,d. The simulation is
for an incident energy of 100~keV, with a total instrumental resolution 
of 150~eV, and a scattering angle of 
110$^\circ$. Both distributions have
two clear peaks, associated with the two carbon isotopes. The 
proton energy is higher for $^{13}$C than for $^{12}$C simply because
the target nucleus is heavier.
Arrows indicate the kinetic energy
of protons that scatter from $^{12}$C and $^{13}$C at the 
surface (the `surface channel').
The $^{13}$C peak for model (d) is lower in energy than that for
model (c), reflecting the fact that the $^{13}$C graphene layer
in (d) is underneath a $^{12}$C overlayer.  In addition, the $^{12}$C peak
for model (d) has two components.  The larger peak,
close to 80~keV, is due to scattering
from the $^{12}$C graphene layer at the surface, while the broader 
peak at lower energy is due scattering from carbon in SiC.

LEEM imaging during Si$^{13}$C epitaxy shows that the 
structure depicted in \fig{model}a can be grown on both SiC(0001) and
SiC(000\=1). After epi-layer growth, monolayer (ML) graphene
layers (\fig{model}b) were formed
by raising the temperature to 
1270\C\ and slowly reducing the
background  pressure of disilane
while the surface was imaged~\cite{Tro09}.  When a complete layer
graphene formed, the
sample temperature was quickly reduced to prevent further
decomposition.
After graphene formation,
the samples were transferred (through air) to the MEIS system.

\begin{figure}[h]
\includegraphics[width=6in]{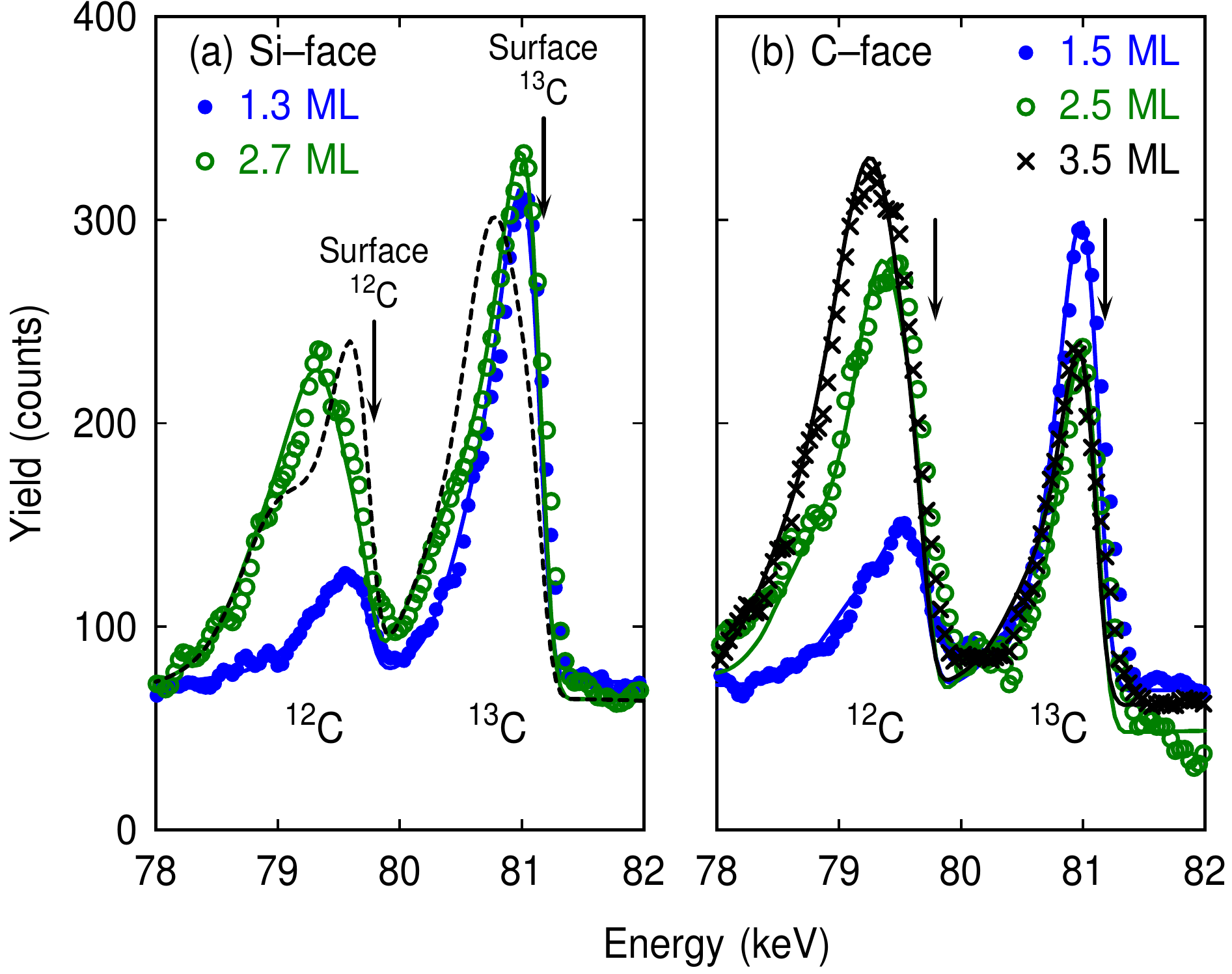}
\caption{100~keV MEIS energy distributions recorded for graphene grown on
(a) SiC(0001) and (b) SiC(000\=1).  Symbols are measured
data, and curves are simulations results.  The color indicates
the average graphene thickness. Blue, green, and
black correspond to approximately 1.5,
2.5, and 3.5 ML of graphene respectively.  The black
dashed curve in (a) indicates a model for 2.7~ML
of graphene in which
the $^{13}$C graphene is located
under the $^{12}$C graphene rather than 
above it (see
text).  This model clearly does not describe the measured
spectrum for 2.7~ML (open green circles).
}
\label{meis}
\end{figure}

Selected data from graphene layers grown on both SiC(0001)
and SiC(0001) are shown in \fig{meis}.
The filled symbols in \fig{meis}a correspond to a sample
with 1.3~ML of graphene.
A convenient parameterization of the
$^{13}$C content is given by 
$F = N_{13} / (N_{12}+N_{13})$,
where $N_{12}$ and $N_{13}$ are the numbers of $^{12}$C and
$^{13}$C atoms in the volume of interest.
For the graphene film $F=0.80$, indicating
small but significant intermixing
during decomposition at 1270\C.
Most likely,
$^{12}$C is incorporated in the graphene due to the
formation of pits during the decomposition~\cite{Han08}.
Pits expose the underlying SiC,
which can then decompose and contribute $^{12}$C to the graphene layer.
The presence of some $^{12}$C in the graphene layer can also result 
from imperfect `reverse' step flow during decomposition. If the 
final step structure is not identical to the structure before 
Si$^{13}$C growth (e.g.\ \fig{afm}c), some Si$^{13}$C will remain
and some Si$^{12}$C
will decompose, contributing $^{12}$C to the graphene.
%
%
%
%
%
%
%
%
Thicker graphene films were produced using a two-step process.
First, $^{13}$C-rich graphene monolayers were formed using the 
method described above: growth of about
three bilayers of epitaxial Si$^{13}$C at 1200\C\
followed by controlled
decomposition at 1270\C\ to form the initial
graphene layer.  Next, additional graphene layers
were formed by annealing for 3' at 1450\C.
MEIS analysis of these films shows that the graphene film 
is indeed thicker.  For example, for the 
film shown in (\fig{meis}a, open symbols), the graphene thickness was 2.7
layers (with the equivalent of
1.7 layers of $^{13}$C and 1.0 layer of $^{12}$C).
The qualitative result is
clear in the raw data shown in \fig{meis}a.  Compared to the 
1.3~ML film, the $^{12}$C peak for the 2.7~ML film is larger, and
the centroid
is shifted to lower energy.  The shift to lower energy indicates
that the bulk of the
$^{12}$C graphene is located further from the surface.
The $^{13}$C peak
has roughly the same intensity and is not shifted in energy.
These observations suggest that the thicker film
contains more $^{12}$C graphene,
but that the additional graphene is located 
below the surface. Quantitative analysis supports
this conclusion: the top half of the
film is $^{13}$C-rich 
($F=0.80$)
while the lower half is
$^{12}$C-rich
($F=0.43$).  This direct
measurement shows that the growth mode
of graphene on SiC(0001) corresponds to \fig{model}c. New
graphene layers form underneath pre-exisiting layers, 
as proposed by \etal{Emtsev}~\cite{Emt08}.
In addition, the MEIS analysis shows that
there is very little bulk C diffusion during 
the brief (minutes) annealing at 1450\C.
Finally, the black dashed line in \fig{meis}a shows the simulated
spectrum for a graphene film with an \emph{inverted} isotope
depth distribution:  $^{12}$C-rich in the top half 
($F=0.43$) and 
$^{13}$C-rich  in the lower half
($F=0.80$).  The disagreement with the 
measured data is striking, indicating the sensitivity of the
MEIS analysis to the isotopic composition.

On the SiC(000\=1) surface, graphene growth would appear
to be quite different.  In contrast to SiC(0001), graphene
grown the C-face is not locked azimuthally to the substrate.
The domains size measured in our experiments is 
generally smaller, and both the
graphene growth rate and nucleation
rate are significantly higher than on SiC(0001).
We performed
MEIS experiments in order to determine if the growth mode
is fundamentally different from that measured for SiC(0001).

Graphene layers of varying thickness were formed on SiC(000\=1)
using a similar procedure to that described above for SiC(0001).
The only significant difference was that, due to the higher
growth rate on SiC(000\=1), thicker graphene films
were formed at 1325\C\ rather than 1450\C.  The MEIS analysis
of 1.5, 2.5 and 3.5~ML films is shown in \fig{meis}b.  In all
cases, the thickness of the epitaxial Si$^{13}$C layer
intially grown corresponded
to about 1.4~ML of graphene.
%
%
%
%
%
%
%
%
For the 1.5 ML film, 3/4 of the $^{13}$C remained in the graphene
layer ($F=0.76$), indicating that the intermixing during the initial
graphene formation is similar to what was observed
for SiC(0001).  For the thicker films, several qualitative 
observations can be made.  First, the area of
the $^{12}$C peak clearly increases with annealing time,
indicating more graphene.  Conversely, 
the area of the $^{13}$C is essentially constant
(within the sample-to-sample variation in the Si$^{13}$C thickness).
Furthermore, the $^{13}$C peak does not shift to lower energy, indicating
that, in all cases, the $^{13}$C graphene is located mainly
at the surface.
Quantitative modeling confirms this view.
For the 2.5~ML film, $F=0.61$ in the top half of the film, while
in the lower half, $F=0.17$. Clearly most of the $^{13}$C remains
at the surface.  For the 3.5~ML the results are similar: $F=0.61$ in
the top third of the film, while $F=0.20$ in the bottom 2/3 of the
graphene film.  Taken together, these results show that when
SiC(000\=1) is annealed at 1325\C,
additional
graphene layers form underneath the initial ($^{13}$C-rich) graphene
layer formed at 1200\C.  That is, the graphene growth mode is essentially
the same on SiC(0001) and SiC(000\=1), despite the stark contrast
in the crystal quality of the graphene layers on SiC(0001) and SiC(000\=1).

In summary, we have directly measured the growth mode of graphene
on SiC(0001) and SiC(000\=1) during surface decomposition
at high-temperature.  Ultra-thin Si$^{13}$C epitaxial `marker layers'
were growth on both surfaces.  Following graphene formation, MEIS was
used to measure the depth distribution of $^{13}$C and $^{12}$C.
At both surfaces (Si-face and C-face), we find
that $^{13}$C is located primarily
in the outmost graphene 
layers.  That is,
despite very different graphene crystallography on the (0001) and
(000\=1) surfaces, the graphene growth mode is the same: new graphene
layers form underneath existing ones.

This work was supported by DARPA under
Contract No.\ FA8650Ð08-C-7838 through the CERA program.  


\begin{thebibliography}{15}%
\makeatletter
\providecommand \@ifxundefined [1]{%
 \@ifx{#1\undefined}
}%
\providecommand \@ifnum [1]{%
 \ifnum #1\expandafter \@firstoftwo
 \else \expandafter \@secondoftwo
 \fi
}%
\providecommand \@ifx [1]{%
 \ifx #1\expandafter \@firstoftwo
 \else \expandafter \@secondoftwo
 \fi
}%
\providecommand \natexlab [1]{#1}%
\providecommand \enquote  [1]{``#1''}%
\providecommand \bibnamefont  [1]{#1}%
\providecommand \bibfnamefont [1]{#1}%
\providecommand \citenamefont [1]{#1}%
\providecommand \href@noop [0]{\@secondoftwo}%
\providecommand \href [0]{\begingroup \@sanitize@url \@href}%
\providecommand \@href[1]{\@@startlink{#1}\@@href}%
\providecommand \@@href[1]{\endgroup#1\@@endlink}%
\providecommand \@sanitize@url [0]{\catcode `\\12\catcode `\$12\catcode
  `\&12\catcode `\#12\catcode `\^12\catcode `\_12\catcode `\%12\relax}%
\providecommand \@@startlink[1]{}%
\providecommand \@@endlink[0]{}%
\providecommand \url  [0]{\begingroup\@sanitize@url \@url }%
\providecommand \@url [1]{\endgroup\@href {#1}{\urlprefix }}%
\providecommand \urlprefix  [0]{URL }%
\providecommand \Eprint [0]{\href }%
\providecommand \doibase [0]{http://dx.doi.org/}%
\providecommand \selectlanguage [0]{\@gobble}%
\providecommand \bibinfo  [0]{\@secondoftwo}%
\providecommand \bibfield  [0]{\@secondoftwo}%
\providecommand \translation [1]{[#1]}%
\providecommand \BibitemOpen [0]{}%
\providecommand \bibitemStop [0]{}%
\providecommand \bibitemNoStop [0]{.\EOS\space}%
\providecommand \EOS [0]{\spacefactor3000\relax}%
\providecommand \BibitemShut  [1]{\csname bibitem#1\endcsname}%
\let\auto@bib@innerbib\@empty
\bibitem [{\citenamefont {Novoselov}\ \emph {et~al.}(2005)\citenamefont
  {Novoselov}, \citenamefont {Geim}, \citenamefont {Morozov}, \citenamefont
  {Jiang}, \citenamefont {Katsnelson}, \citenamefont {Grigorieva},
  \citenamefont {Dubonos},\ and\ \citenamefont {Firsov}}]{Nov05}%
  \BibitemOpen
  \bibfield  {author} {\bibinfo {author} {\bibfnamefont {K.~S.}\ \bibnamefont
  {Novoselov}}, \bibinfo {author} {\bibfnamefont {A.~K.}\ \bibnamefont {Geim}},
  \bibinfo {author} {\bibfnamefont {S.~V.}\ \bibnamefont {Morozov}}, \bibinfo
  {author} {\bibfnamefont {D.}~\bibnamefont {Jiang}}, \bibinfo {author}
  {\bibfnamefont {M.~I.}\ \bibnamefont {Katsnelson}}, \bibinfo {author}
  {\bibfnamefont {I.~V.}\ \bibnamefont {Grigorieva}}, \bibinfo {author}
  {\bibfnamefont {S.~V.}\ \bibnamefont {Dubonos}}, \ and\ \bibinfo {author}
  {\bibfnamefont {A.~A.}\ \bibnamefont {Firsov}},\ }\href@noop {} {\bibfield
  {journal} {\bibinfo  {journal} {Nature}\ }\textbf {\bibinfo {volume} {438}},\
  \bibinfo {pages} {197} (\bibinfo {year} {2005})}\BibitemShut {NoStop}%
\bibitem [{\citenamefont {Zhang}\ \emph {et~al.}(2005)\citenamefont {Zhang},
  \citenamefont {Tan}, \citenamefont {Stormer},\ and\ \citenamefont
  {Kim}}]{Zha05}%
  \BibitemOpen
  \bibfield  {author} {\bibinfo {author} {\bibfnamefont {Y.}~\bibnamefont
  {Zhang}}, \bibinfo {author} {\bibfnamefont {Y.-W.}\ \bibnamefont {Tan}},
  \bibinfo {author} {\bibfnamefont {H.~L.}\ \bibnamefont {Stormer}}, \ and\
  \bibinfo {author} {\bibfnamefont {P.}~\bibnamefont {Kim}},\ }\href@noop {}
  {\bibfield  {journal} {\bibinfo  {journal} {Nature}\ }\textbf {\bibinfo
  {volume} {438}},\ \bibinfo {pages} {201} (\bibinfo {year}
  {2005})}\BibitemShut {NoStop}%
\bibitem [{\citenamefont {Berger}\ \emph {et~al.}(2006)\citenamefont {Berger},
  \citenamefont {Son}, \citenamefont {Li}, \citenamefont {Wu}, \citenamefont
  {Brown}, \citenamefont {Naud}, \citenamefont {Mayou}, \citenamefont {Li},
  \citenamefont {Hass}, \citenamefont {Marchenkov}, \citenamefont {Conrad},
  \citenamefont {First},\ and\ \citenamefont {{de Heer}}}]{Ber06}%
  \BibitemOpen
  \bibfield  {author} {\bibinfo {author} {\bibfnamefont {C.}~\bibnamefont
  {Berger}}, \bibinfo {author} {\bibfnamefont {Z.}~\bibnamefont {Son}},
  \bibinfo {author} {\bibfnamefont {X.}~\bibnamefont {Li}}, \bibinfo {author}
  {\bibfnamefont {X.}~\bibnamefont {Wu}}, \bibinfo {author} {\bibfnamefont
  {N.}~\bibnamefont {Brown}}, \bibinfo {author} {\bibfnamefont
  {C.}~\bibnamefont {Naud}}, \bibinfo {author} {\bibfnamefont {D.}~\bibnamefont
  {Mayou}}, \bibinfo {author} {\bibfnamefont {T.}~\bibnamefont {Li}}, \bibinfo
  {author} {\bibfnamefont {J.}~\bibnamefont {Hass}}, \bibinfo {author}
  {\bibfnamefont {A.~N.}\ \bibnamefont {Marchenkov}}, \bibinfo {author}
  {\bibfnamefont {E.~H.}\ \bibnamefont {Conrad}}, \bibinfo {author}
  {\bibfnamefont {P.~N.}\ \bibnamefont {First}}, \ and\ \bibinfo {author}
  {\bibfnamefont {W.~A.}\ \bibnamefont {{de Heer}}},\ }\href@noop {} {\bibfield
   {journal} {\bibinfo  {journal} {Science}\ }\textbf {\bibinfo {volume}
  {312}},\ \bibinfo {pages} {1191} (\bibinfo {year} {2006})}\BibitemShut
  {NoStop}%
\bibitem [{\citenamefont {Forbeaux}\ \emph {et~al.}(1998)\citenamefont
  {Forbeaux}, \citenamefont {Themlin},\ and\ \citenamefont {Debever}}]{For98}%
  \BibitemOpen
  \bibfield  {author} {\bibinfo {author} {\bibfnamefont {I.}~\bibnamefont
  {Forbeaux}}, \bibinfo {author} {\bibfnamefont {J.-M.}\ \bibnamefont
  {Themlin}}, \ and\ \bibinfo {author} {\bibfnamefont {J.-M.}\ \bibnamefont
  {Debever}},\ }\href@noop {} {\bibfield  {journal} {\bibinfo  {journal} {Phys.
  Rev. B}\ }\textbf {\bibinfo {volume} {58}},\ \bibinfo {pages} {16396}
  (\bibinfo {year} {1998})}\BibitemShut {NoStop}%
\bibitem [{\citenamefont {Hass}\ \emph {et~al.}(2006)\citenamefont {Hass},
  \citenamefont {Feng}, \citenamefont {Li}, \citenamefont {Li}, \citenamefont
  {Zong}, \citenamefont {de~Heer}, \citenamefont {First}, \citenamefont
  {Conrad}, \citenamefont {Jeffrey},\ and\ \citenamefont {Berger}}]{Has06}%
  \BibitemOpen
  \bibfield  {author} {\bibinfo {author} {\bibfnamefont {J.}~\bibnamefont
  {Hass}}, \bibinfo {author} {\bibfnamefont {R.}~\bibnamefont {Feng}}, \bibinfo
  {author} {\bibfnamefont {T.}~\bibnamefont {Li}}, \bibinfo {author}
  {\bibfnamefont {X.}~\bibnamefont {Li}}, \bibinfo {author} {\bibfnamefont
  {Z.}~\bibnamefont {Zong}}, \bibinfo {author} {\bibfnamefont {W.~A.}\
  \bibnamefont {de~Heer}}, \bibinfo {author} {\bibfnamefont {P.~N.}\
  \bibnamefont {First}}, \bibinfo {author} {\bibfnamefont {E.~H.}\ \bibnamefont
  {Conrad}}, \bibinfo {author} {\bibfnamefont {C.~A.}\ \bibnamefont {Jeffrey}},
  \ and\ \bibinfo {author} {\bibfnamefont {C.}~\bibnamefont {Berger}},\
  }\href@noop {} {\bibfield  {journal} {\bibinfo  {journal} {Applied Physics
  Letters}\ }\textbf {\bibinfo {volume} {89}},\ \bibinfo {eid} {143106}
  (\bibinfo {year} {2006})}\BibitemShut {NoStop}%
\bibitem [{\citenamefont {Hass}\ \emph {et~al.}(2008)\citenamefont {Hass},
  \citenamefont {Varchon}, \citenamefont {Mill\'an-Otoya}, \citenamefont
  {Sprinkle}, \citenamefont {Sharma}, \citenamefont {de~Heer}, \citenamefont
  {Berger}, \citenamefont {First}, \citenamefont {Magaud},\ and\ \citenamefont
  {Conrad}}]{Has08}%
  \BibitemOpen
  \bibfield  {author} {\bibinfo {author} {\bibfnamefont {J.}~\bibnamefont
  {Hass}}, \bibinfo {author} {\bibfnamefont {F.}~\bibnamefont {Varchon}},
  \bibinfo {author} {\bibfnamefont {J.~E.}\ \bibnamefont {Mill\'an-Otoya}},
  \bibinfo {author} {\bibfnamefont {M.}~\bibnamefont {Sprinkle}}, \bibinfo
  {author} {\bibfnamefont {N.}~\bibnamefont {Sharma}}, \bibinfo {author}
  {\bibfnamefont {W.~A.}\ \bibnamefont {de~Heer}}, \bibinfo {author}
  {\bibfnamefont {C.}~\bibnamefont {Berger}}, \bibinfo {author} {\bibfnamefont
  {P.~N.}\ \bibnamefont {First}}, \bibinfo {author} {\bibfnamefont
  {L.}~\bibnamefont {Magaud}}, \ and\ \bibinfo {author} {\bibfnamefont {E.~H.}\
  \bibnamefont {Conrad}},\ }\href@noop {} {\bibfield  {journal} {\bibinfo
  {journal} {Phys. Rev. Lett.}\ }\textbf {\bibinfo {volume} {100}},\ \bibinfo
  {pages} {125504} (\bibinfo {year} {2008})}\BibitemShut {NoStop}%
\bibitem [{\citenamefont {Emtsev}\ \emph {et~al.}(2008)\citenamefont {Emtsev},
  \citenamefont {Speck}, \citenamefont {Seyller}, \citenamefont {Ley},\ and\
  \citenamefont {Riley}}]{Emt08}%
  \BibitemOpen
  \bibfield  {author} {\bibinfo {author} {\bibfnamefont {K.~V.}\ \bibnamefont
  {Emtsev}}, \bibinfo {author} {\bibfnamefont {F.}~\bibnamefont {Speck}},
  \bibinfo {author} {\bibfnamefont {T.}~\bibnamefont {Seyller}}, \bibinfo
  {author} {\bibfnamefont {L.}~\bibnamefont {Ley}}, \ and\ \bibinfo {author}
  {\bibfnamefont {J.~D.}\ \bibnamefont {Riley}},\ }\href@noop {} {\bibfield
  {journal} {\bibinfo  {journal} {Phys. Rev. B}\ }\textbf {\bibinfo {volume}
  {77}},\ \bibinfo {pages} {155303} (\bibinfo {year} {2008})}\BibitemShut
  {NoStop}%
\bibitem [{\citenamefont {{van~Bommel}}\ \emph {et~al.}(1975)\citenamefont
  {{van~Bommel}}, \citenamefont {Crombeen},\ and\ \citenamefont
  {{van~Tooren}}}]{bom75}%
  \BibitemOpen
  \bibfield  {author} {\bibinfo {author} {\bibfnamefont {A.}~\bibnamefont
  {{van~Bommel}}}, \bibinfo {author} {\bibfnamefont {J.}~\bibnamefont
  {Crombeen}}, \ and\ \bibinfo {author} {\bibfnamefont {A.}~\bibnamefont
  {{van~Tooren}}},\ }\href@noop {} {\bibfield  {journal} {\bibinfo  {journal}
  {Surf.\ Sci.}\ }\textbf {\bibinfo {volume} {48}},\ \bibinfo {pages} {463}
  (\bibinfo {year} {1975})}\BibitemShut {NoStop}%
\bibitem [{\citenamefont {{van der Veen}}(1985)}]{van85}%
  \BibitemOpen
  \bibfield  {author} {\bibinfo {author} {\bibfnamefont {J.~F.}\ \bibnamefont
  {{van der Veen}}},\ }\href@noop {} {\bibfield  {journal} {\bibinfo  {journal}
  {Surf.\ Sci.\ Rep.}\ }\textbf {\bibinfo {volume} {5}},\ \bibinfo {pages}
  {199} (\bibinfo {year} {1985})}\BibitemShut {NoStop}%
\bibitem [{\citenamefont {Gusev}\ \emph {et~al.}(1995)\citenamefont {Gusev},
  \citenamefont {Lu}, \citenamefont {Gustafsson},\ and\ \citenamefont
  {Garfunkel}}]{Gus95}%
  \BibitemOpen
  \bibfield  {author} {\bibinfo {author} {\bibfnamefont {E.~P.}\ \bibnamefont
  {Gusev}}, \bibinfo {author} {\bibfnamefont {H.~C.}\ \bibnamefont {Lu}},
  \bibinfo {author} {\bibfnamefont {T.}~\bibnamefont {Gustafsson}}, \ and\
  \bibinfo {author} {\bibfnamefont {E.}~\bibnamefont {Garfunkel}},\ }\href@noop
  {} {\bibfield  {journal} {\bibinfo  {journal} {Phys. Rev. B}\ }\textbf
  {\bibinfo {volume} {52}},\ \bibinfo {pages} {1759} (\bibinfo {year}
  {1995})}\BibitemShut {NoStop}%
\bibitem [{\citenamefont {Tromp}\ and\ \citenamefont {Hannon}(2009)}]{Tro09}%
  \BibitemOpen
  \bibfield  {author} {\bibinfo {author} {\bibfnamefont {R.~M.}\ \bibnamefont
  {Tromp}}\ and\ \bibinfo {author} {\bibfnamefont {J.~B.}\ \bibnamefont
  {Hannon}},\ }\href@noop {} {\bibfield  {journal} {\bibinfo  {journal} {Phys.
  Rev. Lett.}\ }\textbf {\bibinfo {volume} {102}},\ \bibinfo {pages} {106104}
  (\bibinfo {year} {2009})}\BibitemShut {NoStop}%
\bibitem [{\citenamefont {Sun}\ \emph {et~al.}(2011)\citenamefont {Sun},
  \citenamefont {Hannon}, \citenamefont {Tromp},\ and\ \citenamefont
  {Poh.}}]{Sun11}%
  \BibitemOpen
  \bibfield  {author} {\bibinfo {author} {\bibfnamefont {J.}~\bibnamefont
  {Sun}}, \bibinfo {author} {\bibfnamefont {J.}~\bibnamefont {Hannon}},
  \bibinfo {author} {\bibfnamefont {R.}~\bibnamefont {Tromp}}, \ and\ \bibinfo
  {author} {\bibfnamefont {K.}~\bibnamefont {Poh.}},\ }\href@noop {} {\bibfield
   {journal} {\bibinfo  {journal} {IBM J.\ Res.\ Develop.}\ }\textbf {\bibinfo
  {volume} {NA}},\ \bibinfo {pages} {NA} (\bibinfo {year} {2011})}\BibitemShut
  {NoStop}%
\bibitem [{\citenamefont {Hannon}\ \emph {et~al.}(2009)\citenamefont {Hannon},
  \citenamefont {Tromp}, \citenamefont {Medhekar},\ and\ \citenamefont
  {Shenoy}}]{Han09}%
  \BibitemOpen
  \bibfield  {author} {\bibinfo {author} {\bibfnamefont {J.~B.}\ \bibnamefont
  {Hannon}}, \bibinfo {author} {\bibfnamefont {R.~M.}\ \bibnamefont {Tromp}},
  \bibinfo {author} {\bibfnamefont {N.~V.}\ \bibnamefont {Medhekar}}, \ and\
  \bibinfo {author} {\bibfnamefont {V.~B.}\ \bibnamefont {Shenoy}},\
  }\href@noop {} {\bibfield  {journal} {\bibinfo  {journal} {Phys. Rev. Lett.}\
  }\textbf {\bibinfo {volume} {103}},\ \bibinfo {pages} {256101} (\bibinfo
  {year} {2009})}\BibitemShut {NoStop}%
\bibitem [{\citenamefont {Hannon}\ and\ \citenamefont {Tromp}(2008)}]{Han08}%
  \BibitemOpen
  \bibfield  {author} {\bibinfo {author} {\bibfnamefont {J.~B.}\ \bibnamefont
  {Hannon}}\ and\ \bibinfo {author} {\bibfnamefont {R.~M.}\ \bibnamefont
  {Tromp}},\ }\href@noop {} {\bibfield  {journal} {\bibinfo  {journal} {Phys.\
  Rev.\ B}\ }\textbf {\bibinfo {volume} {77}},\ \bibinfo {eid} {241404}
  (\bibinfo {year} {2008})}\BibitemShut {NoStop}%
\bibitem [{\citenamefont {Copel}\ \emph {et~al.}(2011)\citenamefont {Copel},
  \citenamefont {Oida}, \citenamefont {Kasry}, \citenamefont {Bol},
  \citenamefont {Hannon},\ and\ \citenamefont {Tromp}}]{Cop11}%
  \BibitemOpen
  \bibfield  {author} {\bibinfo {author} {\bibfnamefont {M.}~\bibnamefont
  {Copel}}, \bibinfo {author} {\bibfnamefont {S.}~\bibnamefont {Oida}},
  \bibinfo {author} {\bibfnamefont {A.}~\bibnamefont {Kasry}}, \bibinfo
  {author} {\bibfnamefont {A.~A.}\ \bibnamefont {Bol}}, \bibinfo {author}
  {\bibfnamefont {J.~B.}\ \bibnamefont {Hannon}}, \ and\ \bibinfo {author}
  {\bibfnamefont {R.~M.}\ \bibnamefont {Tromp}},\ }\href@noop {} {\bibfield
  {journal} {\bibinfo  {journal} {Applied Physics Letters}\ }\textbf {\bibinfo
  {volume} {98}},\ \bibinfo {pages} {113103} (\bibinfo {year}
  {2011})}\BibitemShut {NoStop}%
\end{thebibliography}

%

\end{document}